\documentclass[%
 reprint,
amsmath,amssymb,
aps,
prl,
]{revtex4-1}

\usepackage{epsfig}
\usepackage{array}
\usepackage{multirow}
\usepackage{amsmath}
\usepackage{graphicx}
\usepackage{dcolumn}
\usepackage{bm}
\usepackage{braket} 

\begin{document}

\title{Spin-conserving and reversing photoemission \\ from the surface states of Bi$_2$Se$_3$ and Au (111)}

\author{Ji Hoon Ryoo}
\author{Cheol-Hwan Park}
\email{cheolhwan@snu.ac.kr}
\affiliation{
Department of Physics, Seoul National University, Seoul 08826, Korea}

\date{\today}

\begin{abstract}
We present a theory based on first-principles calculations
explaining
(i) why the tunability of spin polarizations of photoelectrons from Bi$_2$Se$_3$ (111) depends on the band index and Bloch wavevector of the surface state
and
(ii) why such tunability is absent in
the case of {\it isosymmetric} Au (111).
The results provide not only an explanation for
the recent, puzzling experimental observations
but also a guide toward making highly-tunable spin-polarized
electron sources from topological insulators.
\end{abstract}

\maketitle



Since the beginning of spintronics, constant efforts have been made to generate electrons with a high degree of spin polarization
using transport~\cite{johnson1985interfacial}, optical~\cite{pierce1976photoemission}, and magnetic resonance 
methods~\cite{sarma2000theoretical}.
In particular, optical methods, also known as optical spin orientation,
use polarized-light irradiation. For example, electrons in the valence band of
strained and surface-treated GaAs can be excited by circularly polarized light and emitted with 
$\sim80~\%$ spin polarization~\cite{nakanishi1991large}. GaAs photocathodes are widely used as spin-polarized electron source in low-energy electron microscopy~\cite{bauer2002spin},
in accelerators used in high-energy physics~\cite{alley1995stanford},
etc.

Recently, it has been proposed that topological insulators can serve as a spin-polarized electron source when irradiated with polarized light~\cite{jozwiak2013photoelectron}.
By changing the polarization of light and the direction toward
which photoelectrons are collected,
one can obtain an electron beam which is spin-polarized
in an arbitrary direction, with a 100~\%
degree of spin polarization~\cite{park2012spin}
(the measured degree is over 80~\%~\cite{jozwiak2011widespread}).
On the other hand, the direction of spin polarization
of electrons generated from a strained-GaAs photocathode is
fixed by the surface-normal direction perpendicular
to which the strain is applied.
Moreover, unlike GaAs photocathodes,
in which the photon energy is fixed to $\sim1.5$~eV
by the material band gap,
photocathodes using a topological insulator can
be operated within a wide range of photon energies.
Even if there could be several technological hurdles that should
be overcome, topological insulators are
conceptually new candidates for photocathodes for spintronics.

\begin{figure}
\includegraphics[width=\columnwidth]{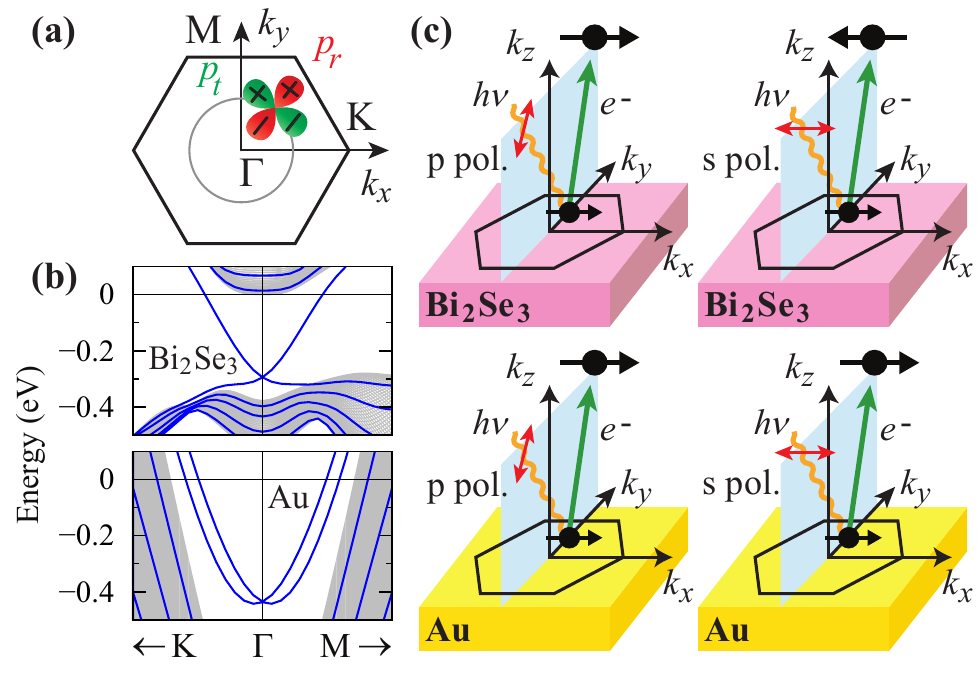}
\caption{
(a) The Brillouin zone of Bi$_2$Se$_3$ and Au surfaces.
The $p_r$ and $p_t$ orbitals of a given Bloch
state specified by the Bloch wavevector
${\bf k}=k\,\left(\cos\phi_{\bf k},\,\sin\phi_{\bf k}\right)$
are defined as $p_r=\cos\phi_{\bf k}\,p_x+\sin\phi_{\bf k}\,p_y$
and $p_t=-\sin\phi_{\bf k}\,p_x+\cos\phi_{\bf k}\,p_y$, respectively, where $p_x$ and $p_y$ are the valence
atomic {\it p} orbitals.
(b) The bandstructures (blue curves)
of the (111) surfaces of Bi$_2$Se$_3$ and Au
along $(0.14,\,0)-(0,\,0)-(0,\,0.14)$ 
in reciprocal space in units of
$2\pi/a$, where $a$ is the lattice parameter.
Projected bulk bands are also shown in gray.
(c) Schematics of the SARPES experimental setup and the results of Ref.~\cite{jozwiak2013photoelectron}.
The horizontal arrows denote the direction of
the spin polarization of the surface electrons
and photoelectrons.}
\label{fig1}
\end{figure}

Despite these recent developments, we still do not understand
the results from some of
the key spin- and angle-resolved photoemission spectroscopy
(SARPES)
experiments on the surface of the Bi$_2$Se$_3$ family of topological insulators, whose space group is R$\bar{3}$m, such as Bi$_2$Se$_3$, Bi$_2$Te$_3$ and Sb$_2$Te$_3$.
\citet{jozwiak2013photoelectron} studied photoelectrons ejected
from the Dirac-cone-like surface band of Bi$_2$Se$_3$
and the Rashba-split surface band of Au
(Fig. \ref{fig1}).
When shone on a Bi$_2$Se$_3$ (111) surface,
p-polarized light generates photoelectrons
whose spin direction is parallel
to that of the surface electrons, while s-polarized light
produces photoelectrons with the opposite
spin~\cite{jozwiak2013photoelectron}.
Since the Bi$_2$Se$_3$ and Au (111) surfaces have the same symmetry, the theoretical analysis predicts
that the gold surface would also exhibit the same photo-induced spin modulation as Bi$_2$Se$_3$ (111)~\cite{park2012spin}.
The SARPES experiment for the gold surface~\cite{jozwiak2013photoelectron}
clearly resolves the two spin-split bands; however, both s- and p-polarized lights produce photoelectrons
with the same spin direction as
that of the initial state in each surface band [Fig. \ref{fig1}(c)].

Also, another experimental study on Bi$_2$Se$_3$ (111) has shown that photoemission
from the upper branch of the surface bands exhibits such photo-induced spin modulation, while that from the lower branch does not~\cite{xie2014orbital}.
In an effort to explain this observation, it was claimed that s-polarized light probes the spinor that couples to $p_r$ orbital, because the electronic states in the lower branch have more $p_r$ character than $p_t$ one~\cite{zhang2013spin}
[for the definition of $p_r$ and $p_t$ orbitals, see Fig.~\ref{fig1}(a)].
However, the spinor being measured is the one coupled to the
orbital interacting with s-polarized light ($p_t$),
and not the one coupled to the dominant $p$ orbital ($p_r$).
Therefore, the experimental observation cannot be understood
from previous theories~\cite{park2012spin,zhang2013spin}.

In summary, we still do not have a good understanding of the photo-induced spin modulation
phenomenon involving the Bi$_2$Se$_3$ family of topological insulators.
In this study, we perform first-principles calculations on
the spin polarization of photoelectrons ejected from
the Bi$_2$Se$_3$ and Au (111) surfaces.
First of all, our results agree with the recent experimental
observations in 
Refs.~\cite{jozwiak2013photoelectron,xie2014orbital}
that were not understood before.
We show that the complicated, material-dependent
coupling between the
spinor part and the orbital part of the wavefunctions
plays a central role in determining the spin polarization
from these surfaces.
We also show that this spinor-orbital coupling in the
wavefunction of Bi$_2$Se$_3$, in particular,
depends heavily on both
the direction and magnitude of the Bloch wavevector;
the pronounced deviation of the spinor-orbital coupling
from the one near the Dirac point is seen in the lower branch
along $\mathrm{\Gamma K}$, where the low-energy
effective theories~\cite{park2012spin,zhang2013spin}
predict that the direction of the
spin polarization of photoelectrons is the opposite
of the experimental observation~\cite{xie2014orbital}.
Our results provide a theoretical background
for developing next-generation
spin-polarized electron sources.


To obtain the spin polarization of photoelectrons, we calculate the matrix elements of
$\mathbf{A}\cdot\mathbf{p}$, where $\mathbf{A}$ is
a vector parallel to
the polarization of light
and $\mathbf{p}$ the momentum operator,
between the initial surface state and the
two (spin-up and spin-down) photoexcited states.
This method of using the dipole transition operator
to account for light-matter interactions
reproduces the measured spin polarization
of photoelectrons ejected from Bi$_2$Se$_3$
quite successfully~\cite{zhu2013layer,zhu2014photoelectron}.
For computational details, see Supplemental Material~\cite{SI}.
To simulate low-energy photoemission
experiments~\cite{jozwiak2013photoelectron,xie2014orbital}, we set the photon energy to 6\,eV.

We denote the incoming direction
of incident photons by
$(-\sin \theta_{\textrm{ph}} \cos \phi_{\textrm{ph}},\, -\sin \theta_{\textrm{ph}} \sin \phi_{\textrm{ph}},\, -\cos \theta_{\textrm{ph}})$
and the outgoing direction of photoelectrons by
$(\sin \theta_{e} \cos \phi_{e},\, \sin \theta_{e} \sin \phi_{e},\, \cos \theta_{e})$.
We focus mainly on two cases:
$\phi_{\textrm{ph}}=\pm 90^{\circ}$ and
$\phi_{e}=\pm 90^{\circ}$ (i.\,e.\,, the in-plane momenta of light and photoelectrons are along $\mathrm{\Gamma M}$)
and $\phi_{\textrm{ph}}=0^{\circ}$ or $180^{\circ}$ and $\phi_{e}=0^{\circ}$ or $180^{\circ}$ (along $\mathrm{\Gamma K}$).

\begin{figure}
\includegraphics[width=\columnwidth]{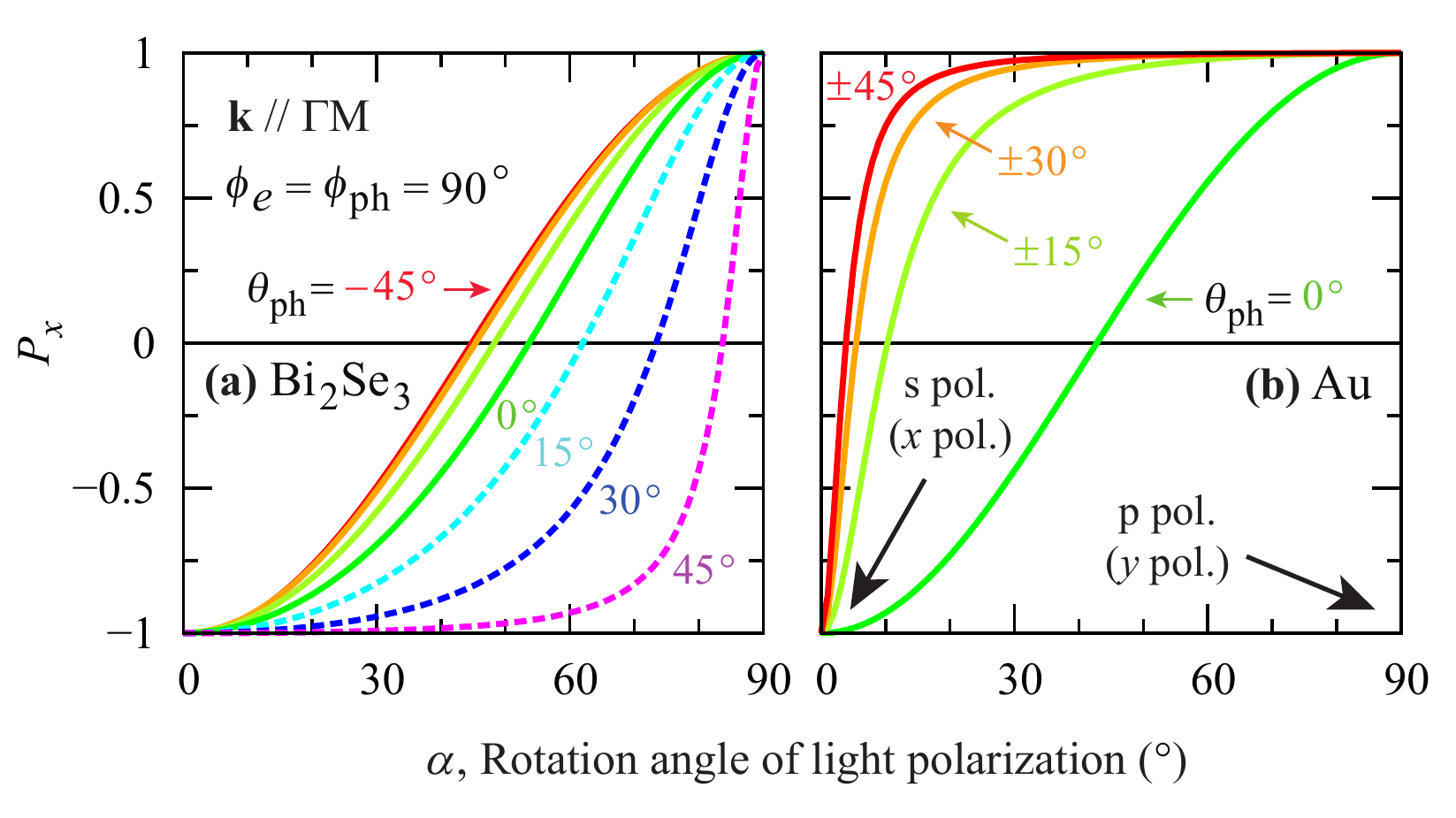}
\caption{The spin polarization along $x$, $P_x$,
of photoelectrons emitted from Bi$_2$Se$_3$ (111) [(a)] and
from Au (111) [(b)].
Note that, for notational convenience,
we have used negative $\theta_{\textrm{ph}}$
to denote cases with $\phi_{\textrm{ph}}=-90^{\circ}$.
The initial surface state is in the upper band
and has $\mathbf{k}=0.01\, (2\pi/a)\, \hat{y}$.}
\label{s2p}
\end{figure}

First, we compare the spin polarization of photoelectrons emitted from the upper band of the surface states
of Bi$_2$Se$_3$ (111) [Fig.~\ref{fig1}(c)] and
of Au (111) [Fig.~\ref{fig1}(d)]
when the incident photons and photoelectrons
both lie in the mirror plane, which is
perpendicular to $\mathrm{\Gamma K}$
($\phi_{\textrm{ph}}=\pm 90^{\circ}$ and $\phi_e=90^{\circ}$).
[For $\phi_e=-90^{\circ}$, similar results are obtained
provided the sign of $\phi_{\textrm{ph}}$ is flipped in
Fig.~\ref{s2p} (not shown)].
We define the spin polarization vector (without $\hbar/2$)
$\mathbf{P}$ of a certain state
as the expectation value of the Pauli spin operators
taken for that state.
Then, due to the mirror symmetry,
(i) $\mathbf{P}$ of any surface state with Bloch wavevector
{\bf k} along $\mathrm{\Gamma M}$ is parallel or antiparallel
to $\mathrm{\Gamma K}$ and
(ii) the p- and s-polarized photons generate photoelectrons
characterized by $\mathbf{P}$ which is 100~\% in magnitude
and is, respectively, parallel to and antiparallel to the
$\mathbf{P}$ of the surface state~\cite{park2012spin}.
This symmetry analysis is in agreement with the
SARPES experimental results on Bi$_2$Se$_3$ (111)~\cite{jozwiak2013photoelectron,zhu2014photoelectron, cao2012coupled}.

Since Bi$_2$Se$_3$ (111) and Au (111) have the same
symmetry, one would naturally expect that the same symmetry analysis
holds for Au (111); however, it was observed that
photoelectrons from the gold surface have the same
spin polarization independent of the direction of {\bf A}~\cite{jozwiak2013photoelectron}.

In order to understand these seemingly contradictory results for Au (111),
we calculate $P_x$ of photoelectrons as a function of the rotation angle $\alpha$
of light polarization (Fig.~\ref{s2p}).
For s- and p-polarized light ($\alpha$ being $0^\circ$ and
$90^\circ$, respectively)
the calculated $P_x$
for both Bi$_2$Se$_3$ (111) and Au (111)
is in accord with the symmetry-based theoretical prediction.
A first-principles study also reported the spin reversal of photoelectrons ejected from Au (111) by s-polarized light and the spin conservation by p-polarized light~\cite{henk2003spin}.
However, the Bi$_2$Se$_3$ and Au surfaces
exhibit differences in the manner $P_x$
changes in between (Fig.~\ref{s2p}).

We first consider the case $\theta_{\textrm{ph}} = 45^{\circ}$
(corresponding to the curves in Fig.~\ref{s2p} with
$\theta_{\textrm{ph}} = \pm45^{\circ}$).
For Bi$_2$Se$_3$, $P_x$-versus-$\alpha$ relations
for $\phi_{\textrm{ph}}=90^{\circ}$ and
for $\phi_{\textrm{ph}}=-90^{\circ}$
(denoted by negative $\theta_{\rm ph}$ in Fig.~\ref{s2p})
are qualitatively different [Fig. \ref{s2p} (a)]:
(i) when $\phi_{\textrm{ph}}=-90^{\circ}$, $P_x$
varies slowly with $\alpha$ from $-1$ to $1$,
changing the sign near $\alpha=45^{\circ}$; (ii) when $\phi_{\textrm{ph}}=90^{\circ}$, $P_x$ remains negative
as long as $\alpha<83^{\circ}$.
On the other hand, for Au,
the dependence of $P_x$ on $\alpha$
for $\phi_{\textrm{ph}}=90^{\circ}$ and that for
$\phi_{\textrm{ph}}=-90^{\circ}$ are essentially the same.
In both cases, $P_x$ changes sharply from $-1$ to nearly $1$
at small $\alpha$ [$P_x=0$ at $\alpha=3^{\circ}$; see Fig.~\ref{s2p}(b)].

This difference between the two materials
on how $P_x$ changes with $\alpha$
originates from the difference in the surface-state wavefunctions.
Among the orbitals constituting the (initial) surface states,
we focus on $p$ orbitals which play a dominant role in
photoemission when the final states have {\it s}-like
characters. This scheme successfully describes the
results from {low-energy} SARPES experiments on
Bi$_2$Se$_3$~\cite{zhu2014photoelectron,sanchez2014photoemission}.

\begin{figure}
\includegraphics[width=0.9\columnwidth]{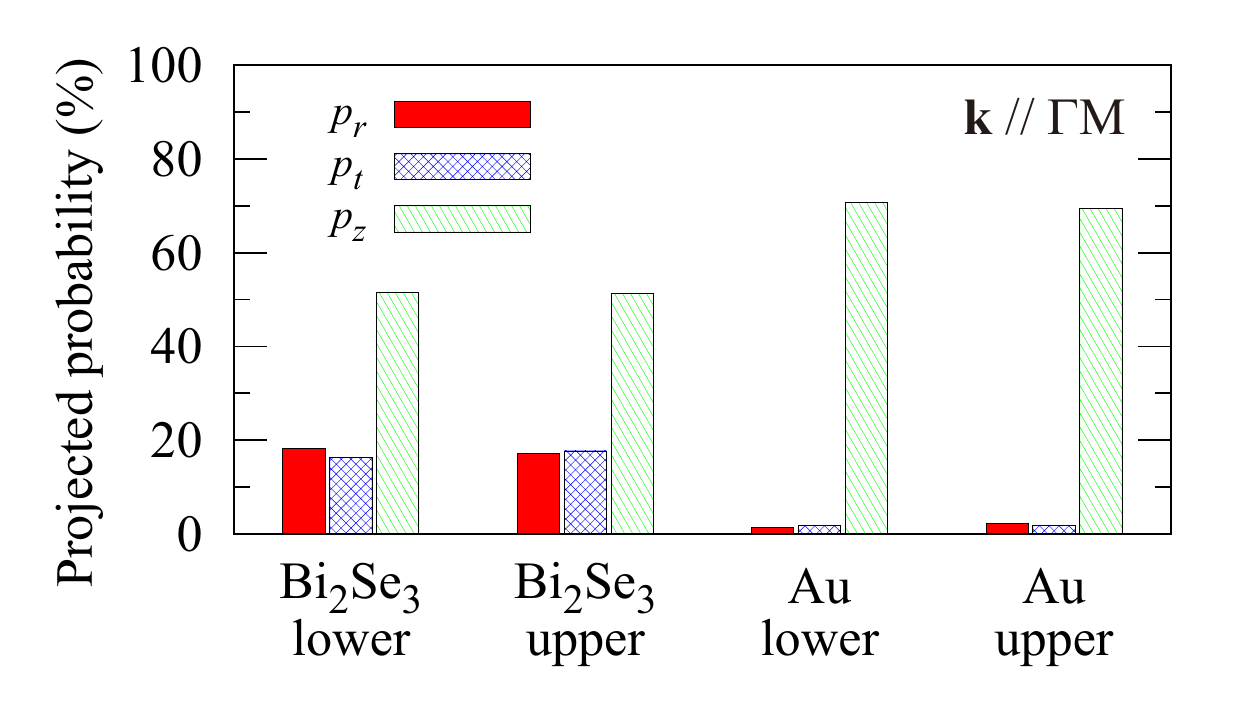}
\caption{
The projected probability (i.\,e.\,, squared amplitude) to each $p$ orbital
of the surface state with $\mathbf{k}=0.01\, (2\pi/a)\,\hat{y}$.
}
\label{tot_spin}
\end{figure}

Figure~\ref{tot_spin} shows squared projections
of the surface states near $\mathrm{\Gamma}$
of Bi$_2$Se$_3$ and Au
to each valence $p$ orbital, summed over atomic sites.
In Bi$_2$Se$_3$ case, the contribution of
in-plane {\it p}-orbitals ($p_r$ and $p_t$)
to the surface states is $35\,\%$,
similar in magnitude to that of $p_z$ orbital (51\,\%).
On the contrary, each in-plane orbital
($p_r$ or $p_t$) of Au contributes less than $2\,\%$
to the surface state of Au (111).
The results on Au (111) are consistent with previous
studies~\cite{lee2012role,ishida2014rashba}.

Although the symmetry analysis indicates that each $p$ orbital comprising the gold surface states couples to spinors
in the same way as in the case of Bi$_2$Se$_3$,
since the surface states of Au has almost no
in-plane $p$-orbital character,
the spin degree of freedom is {\it not}
entangled with the orbital ones.
Therefore, if $A_z$ is finite, even if it is small,
{\bf P} of photoelectrons from the gold surface
is almost completely determined by the spinor coupled to the $p_z$ orbital of the surface state.
Thus, $P_x$ rises sharply
as $\alpha$ deviates from $0^\circ$ [Fig. \ref{s2p}(b)].

We compare $P_x$'s of photoelectrons
associated with $\phi_{\textrm{ph}}=90^{\circ}$ and
that associated with $\phi_{\textrm{ph}}=-90^{\circ}$.
The light polarization vectors for these two cases are
the same except that the signs of their out-of-plane
components are opposite.
Because the spinors attached to in-plane and out-of-plane $p$ orbitals interfere with each other differently in the two cases, the corresponding {\bf P}'s are in principle
different.
This effect is sizable for Bi$_2$Se$_3$ (111)
[Fig.~\ref{s2p}(a)] but is negligible for
Au (111) [Fig.~\ref{s2p}(b)] because, again,
the contribution of in-plane $p$ orbitals to the surface
states of Au (111) is small.

The dependence of $P_x$ on $\theta_{\textrm{ph}}$
(Fig. \ref{s2p}) further illustrates the importance of
the entanglement between the spin and orbital degrees of freedom
in photoemission processes.
When $\theta_{\textrm{ph}}=0^{\circ}$ (i.\,e.\,, normal incidence), $A_z=0$,
and only the in-plane $p$ orbitals are probed.
Therefore, in this case, $P_x$ of photoelectrons from both Bi$_2$Se$_3$ and Au surfaces changes slowly with $\alpha$ from $-1$ to $1$.
When $\theta_{\textrm{ph}}$ increases from $0^{\circ}$ to $15^{\circ}$, $A_z$ becomes finite; therefore, the dependence of $P_x$ of photoelectrons from Au (111) on $\alpha$ significantly changes, becoming similar to that corresponding to $\theta_{\textrm{ph}}=45^{\circ}$.
For Bi$_2$Se$_3$, however, this increase in $\theta_{\textrm{ph}}$
does not have such a huge effect on $P_x$.

From the results of our calculations, we can understand the
hitherto incomprehensible differences in the results of
SARPES experiments on Bi$_2$Se$_3$ and Au surfaces~\cite{jozwiak2013photoelectron}.
If the light with \textit{perfect} s-polarization excites a surface state, the measured {\bf P} must be antiparallel to the
spin polarization of the surface state for both Bi$_2$Se$_3$ and Au.
In real experiments, however, the ``s-polarized'' light
may contain a few percent of the p component
due to
the imperfection of the polarizer,
the inaccuracy in the alignment, or
the inhomogeneity of the surface.
Our calculations [Fig.~\ref{s2p}(b)] suggest that
this small fraction of p-polarized light may determine
the spin polarization of photoelectrons from Au (111), which
explains the experimental result~\cite{jozwiak2013photoelectron}
that s- and p-polarized lights produce photoelectrons with similar {\bf P}'s and that photo-induced spin modulation
is hard to achieve with Au (111).


We now discuss the SARPES configuration
$\phi_{\textrm{ph}}=0^{\circ}$ and $\phi_e=0^{\circ}$, i.\,e.\,,
photons and electrons have the in-plane momenta parallel to $\mathrm{\Gamma K}$. In this case,
no symmetry principle restricts the spin direction of surface electrons or photoelectrons. Nevertheless, when $\mathbf{k}$ of a surface state is small, according to
first-order $\mathbf{k}\cdot\mathbf{p}$ perturbation theory~\cite{zhang2013spin},
the $p_r$ and $p_z$ orbitals in the surface states always couple to the spinor $\ket{\downarrow_t}$
and the $p_t$ orbital to $\ket{\uparrow_t}$
in the upper branch, where $\ket{\uparrow_t}$ and $\ket{\downarrow_t}$ are
the eigenspinors of $\sigma_t=\boldsymbol{\sigma}\cdot(\hat{z}\times\hat{k})$
with eigenvalues 1 and $-1$, respectively.
(The three {\it p} orbitals couple to the opposite spinors in the lower branch.)
Therefore, for a small $k$, {\bf P} of the photoelectrons generated by
p- and s- polarized lights are parallel to and antiparallel to the {\bf P} of the surface state,
respectively.

\begin{figure}
\includegraphics[width=\columnwidth]{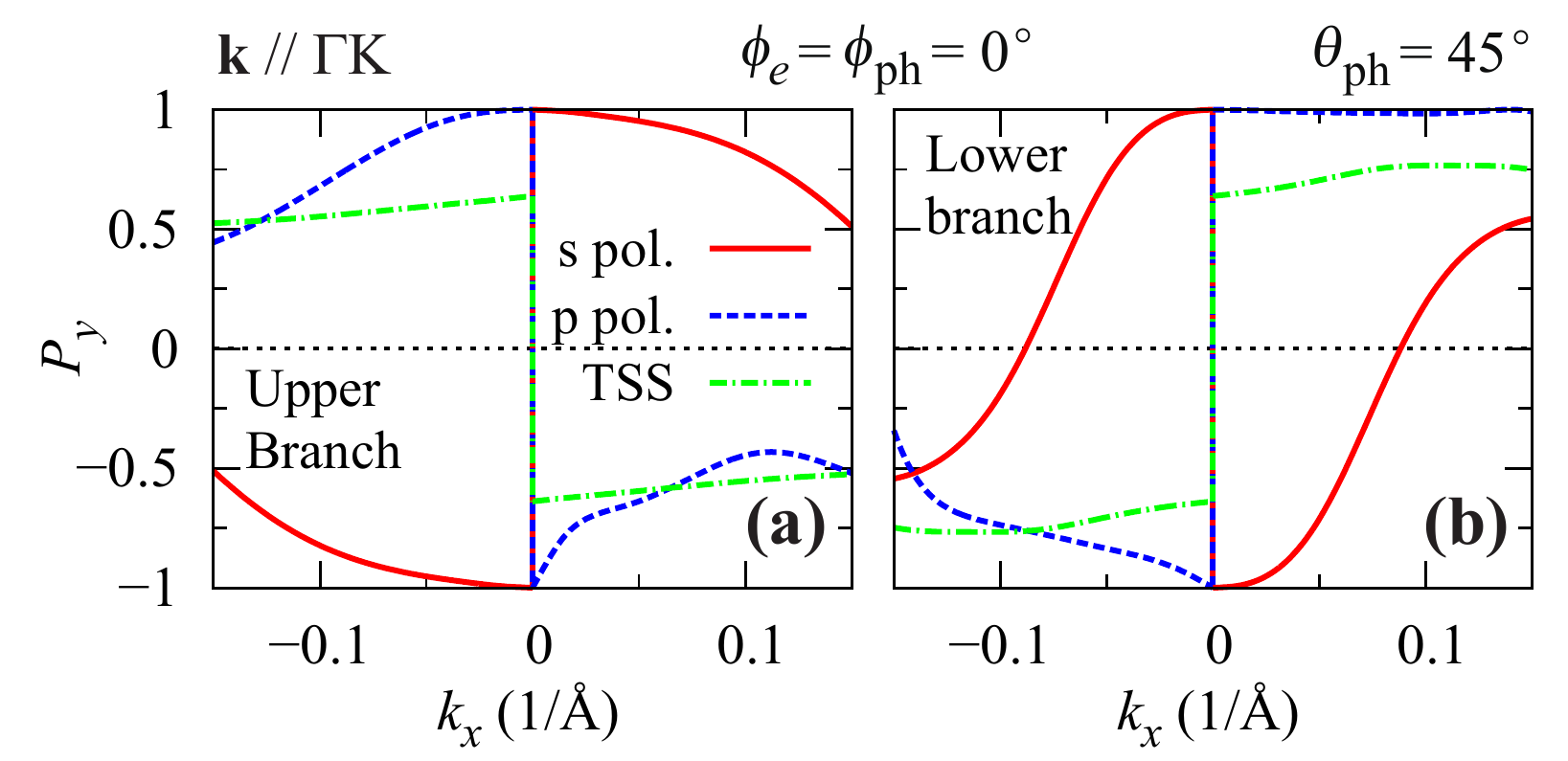}
\caption{
The spin polarization of photoelectrons along $y$, $P_y$,
emitted from Bi$_2$Se$_3$ (111) surface states with $\mathbf{k}$ along $\mathrm{\Gamma K}$.
The dash-dotted or green curve shows $P_y$ of the initial topological surface state (TSS).
\label{ncomms}
}
\end{figure}

However, first-order $\mathbf{k}\cdot\mathbf{p}$ theory is valid only at a small {\it k}:
the couplings between orbitals and spinors that are forbidden near $\Gamma$
(e.\,g.\,, $p_r$ and $\ket{\downarrow_t}$ or $p_t$ and $\ket{\uparrow_t}$ in the lower branch)
are allowed if second or higher order effects are considered.
These couplings are anisotropic in that if {\bf k} is along $\mathrm{\Gamma M}$
they are strictly forbidden even at a large {\it k}.
For Au (111), these higher-order spin-orbital
entanglement effects are difficult to observe due to the dominance of $p_z$ character in the surface state; however, for Bi$_2$Se$_3$ (111), in the lower branch along $\mathrm{\Gamma K}$,
they significantly affect the photoemission process if {\it k} is not small
(Fig.~\ref{ncomms}).

Figure~\ref{ncomms} shows that, when probing the lower branch with large $k_x$, s-polarized light as well as p-polarized light yields photoelectrons whose $P_y$ (the tangential component of {\bf P}) has the same sign as $P_y$ of the surface state, contrary to the small-$k$ results.
In the case of the upper branch, this stark sign change of the spin polarization is not observed in our calculation. The results at large $k$ are confirmed by recent experiments~\cite{xie2014orbital}.

It was suggested that this lack of photo-induced spin modulation
associated with the surface state in the lower branch
was due to the dominance of $p_r$ orbital in the corresponding surface state
which couples to $\left|\uparrow_t\right>$~\cite{xie2014orbital}.
However, since s-polarized light picks up the spinor coupled to $p_t$ orbital
and not the spinor coupled to the dominant $p$ orbital (i.\,e.\,, $p_r$),
this explanation is not satisfactory.
Instead, we show in the following that the origin of this phenomenon is
the complex spin-orbital coupling in the initial surface state
at large {\it k},
which is absent in the low-energy theory~\cite{park2012spin,zhang2013spin}.

\begin{figure}
\includegraphics[width=0.9\columnwidth]{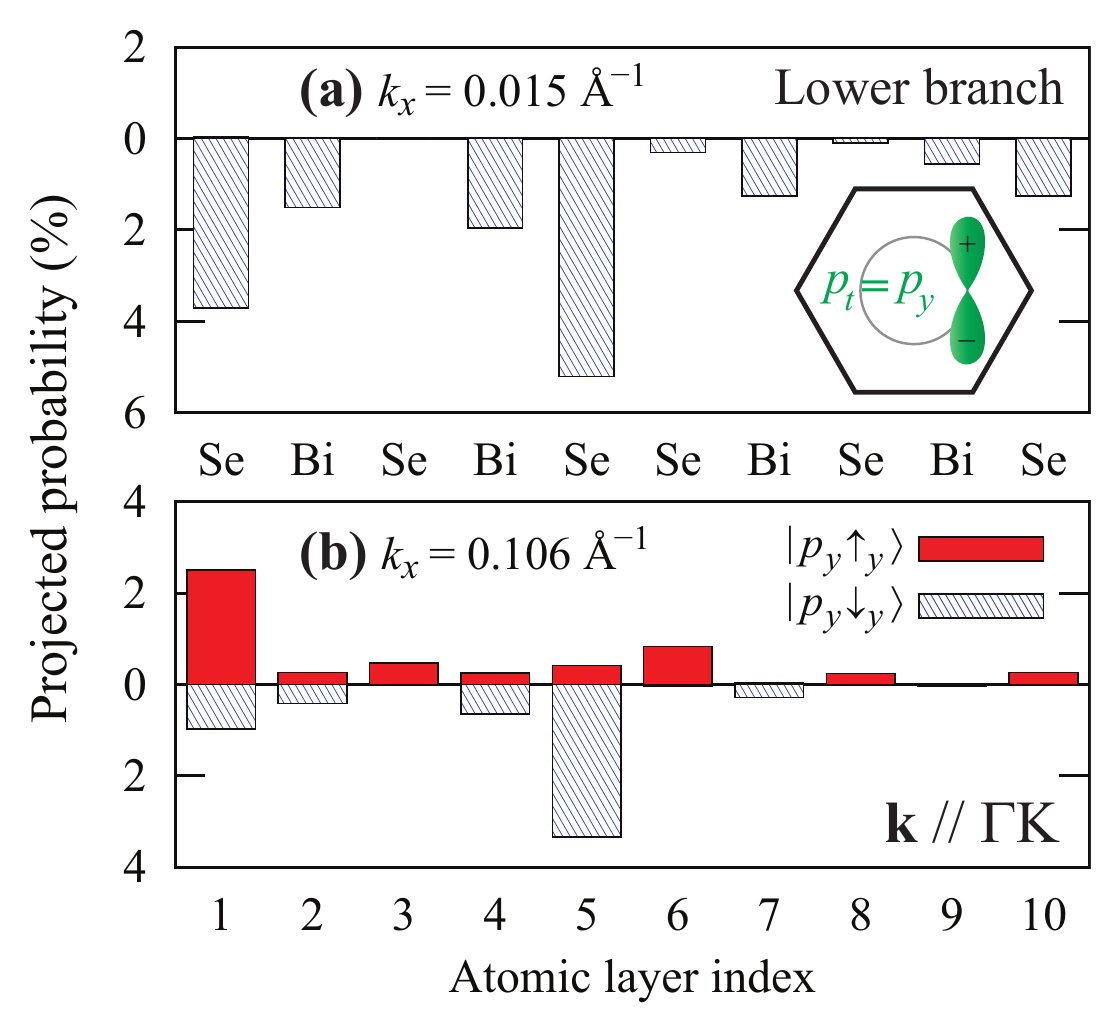}
\caption{Projected probability to the $p_t$ orbital in each atomic layer
of the surface states in the lower branch
at $\mathbf{k} = 0.015\, \mbox{\AA}^{-1}\, \hat{x}$ [(a)]
and at $\mathbf{k} =0.106\, \mbox{\AA}^{-1}\, \hat{x}$ [(b)].
The spin is quantized along $y$.
Atomic layer~1 is the topmost surface layer.}
\label{layer_spin}
\end{figure}

Figure~\ref{layer_spin} shows the extent of contribution of the $p_t$ orbital (which is $p_y$) to the surface states in the lower branch with {\bf k} along $\mathrm{\Gamma K}$,
resolved to each spinor.
Near $\mathrm{\Gamma}$ ($k = 0.015\,\mbox{\AA}^{-1}$), the $p_y$ orbital in each layer
couples exclusively to $\left|\downarrow_y\right>$, as predicted by
first-order $\mathbf{k}\cdot\mathbf{p}$ theory~\cite{zhang2013spin}.
However, when $k = 0.106\,\mbox{\AA}^{-1}$, the coupling of $p_y$ to
$\left|\uparrow_y\right>$ is significant and, especially,
in the case of the topmost layer (which is the most important in photoemission processes),
the projected probability to $\left|\uparrow_y\right>$
is more than twice as high as that to $\left|\downarrow_y\right>$.

\begin{figure}
\includegraphics[width=\columnwidth]{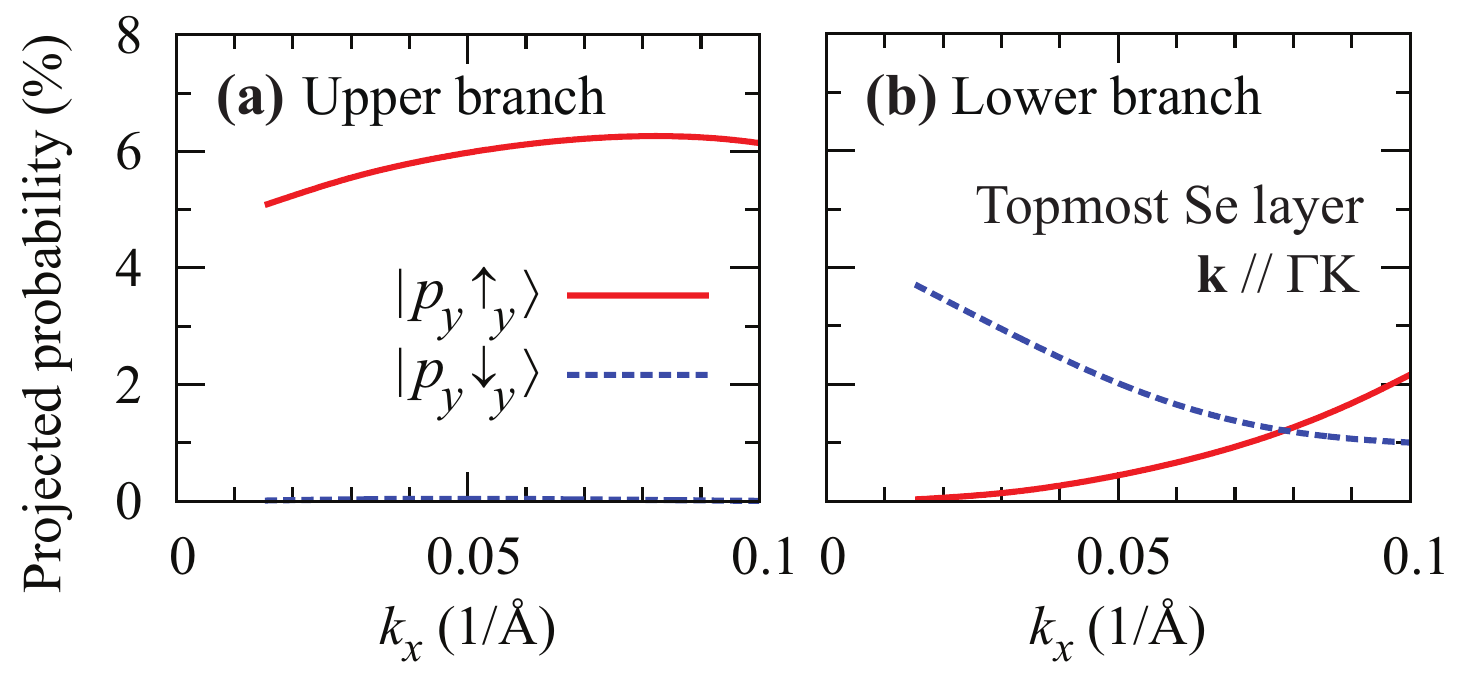}
\caption{
Projected probability to the $p_t$ orbital in the topmost atomic layer of the surface states
with {\bf k} along $\mathrm{\Gamma K}$ as a function of {\it k}.
\label{topmost}
}
\end{figure}

Figure~\ref{topmost} shows the projected probability of the tangential $p$ orbital at the topmost surface layer.
For the surface state in the upper branch,
the contribution from the term $\ket{p_y \downarrow_y}$, albeit not forbidden at large $k$,
is negligible in the range of $k$ considered.
(In fact, this $\ket{p_y \downarrow_y}$ contribution is tiny up to the fourth atomic layers from the surface~\cite{SI}.)
However, for the surface state in the lower branch at large $k$, the contribution of $\ket{p_y \uparrow_y}$ in the topmost layer outweighs that of $\ket{p_y \downarrow_y}$.
(See Supplemental Material~\cite{SI} for the layer-resolved
projection to the three $p$ orbitals.)
This difference explains why, at large $k$,
$P_y$ of the photoelectrons from the upper branch excited by
the s- and p-polarized lights have different signs [Fig.~\ref{ncomms}(a)],
whereas $P_y$ of the photoelectrons from the lower branch have the same sign [Fig.~\ref{ncomms}(b)].

In conclusion, we studied the possibility of modulating the electron spin through photoemission
from the surfaces of Bi$_2$Se$_3$ and Au. We find that both (i) the intricate spin-orbital coupling and (ii) large-$k$ effects are crucial in understanding and predicting
the possibility of photo-induced spin modulation. Not only does our study
provide an explanation of the recent low-energy, spin-dependent photoemission experiments
in a coherent manner, it also establishes a designing principle
for a new kind of spin-polarized electron sources using topological insulators.

We gratefully acknowledge fruitful discussions with X.\,J.\,Zhou on his experimental results
in Ref.~\cite{xie2014orbital} and with Chris Jozwiak and Choongyu Hwang on many aspects of SARPES experiments on Bi$_2$Se$_3$ and Au (111).  This work was supported by Korean NRF-2013R1A1A1076141 funded by MSIP and computational resources were provided by Aspiring Researcher Program through Seoul National University in 2014.

\pagebreak

\begin{center}
\textbf{\large Supplemental Material}
\end{center}

\setcounter{equation}{0}
\setcounter{figure}{0}
\setcounter{table}{0}
\setcounter{page}{1}
\makeatletter
\renewcommand{\theequation}{S\arabic{equation}}
\renewcommand{\thefigure}{S\arabic{figure}}
\renewcommand{\bibnumfmt}[1]{[S#1]}
\renewcommand{\citenumfont}[1]{S#1}

\section{Calculation Details}

We obtained the wavefunctions of surface states
using Quantum Espresso package \cite{S_giannozzi2009quantum}.
We have modeled Bi$_2$Se$_3$ (111) and Au (111) surfaces by
30- and 24-atomic-layer slabs, respectively.
We set the inter-slab distance
to $20\,\mbox{\AA}$ for Bi$_2$Se$_3$ and $30\,\mbox{\AA}$
for Au.
We fully relaxed the lattice parameters and the atomic positions 
taking into account, for Bi$_2$Se$_3$,
van der Waals interactions.
(We have checked that the relaxed structure of {\it bulk}
Bi$_2$Se$_3$ is in very good agreement - less than
1~\% differences in the lattice parameters and
in the inter-quintuple-layer distance - with the
measurement~\cite{S_nakajima1963crystal}.)

To describe ion-electron interactions, we used
fully-relativistic, norm-conserving pseudopotentials.
To account for exchange-correlation interactions,
we used the method of Ref.~\cite{S_perdew1996generalized}
for Bi$_2$Se$_3$ and that of Ref.~\cite{S_perdew2008restoring}
for Au.
The k-point meshes that we used for Bi$_2$Se$_3$~(111)
and for Au~(111) are $13 \times 13 \times 1$
and $12 \times 12 \times 1$, respectively.
The kinetic energy cutoff is set to $60\,\mathrm{Ry}$.

Photoexcited states are described, following
Refs.~\cite{S_zhu2013layer,S_zhu2014photoelectron},
as the Bloch sum of {\it s}-like orbitals inside the crystal.
A photoexcited state $\ket{f}$ can be approximated as, taking into account the phase of
an electron emitted from each atom and inelastic collisions inside the crystal, 
$\ket{f}\approx\ket{\psi}\otimes\ket{\chi}$ with $\braket{\mathbf{r}|\psi}\propto \sum_{{\bf R},\,\alpha}
{e^{z_{\alpha}/2\lambda} e^{i \mathbf{k}_f\cdot(\mathbf{R}+\boldsymbol{\tau}_{\alpha})} \phi_\alpha(\mathbf{r}-\mathbf{R}-\boldsymbol{\tau}_{\alpha})}$
being the orbital part of the wavefunction and $\ket{\chi}$
the spin part (either spin-up or spin-down).
Here, $\mathbf{R}$ denotes the (in-plane) lattice vector,  $\boldsymbol{\tau}_{\alpha}$
the position of an atom $\alpha$ within each unit cell of the slab,
$\lambda$ the inelastic mean free path of electrons, set to $7\,\mbox{\AA}$ \cite{S_zhu2013layer}, and $\phi_\alpha(\mathbf{r}-\mathbf{R}-\boldsymbol{\tau}_{\alpha})$ the orbital localized at each atomic site (see below for details).

The wavevector of a photoelectron $\mathbf{k}_f = (\mathbf{k}_{f,\,\parallel}, k_{f,\,z})$ and the wavevector of the initial surface electron $\mathbf{k}_i$ are related by
$\mathbf{k}_{f,\,\parallel}=\mathbf{k}_i$ and
$k_{f,\,z}=\sqrt{2m(h\nu-E_{\textrm{B}})/\hbar^2-(k_{i,\,x})^2-(k_{i,\,y})^2}$, where $E_{\textrm{B}}$ is the binding energy of the surface state,
$m$ the mass of an electron,
and the photon energy, $h\nu$, is set to $6 \, \textrm{eV}$~\cite{S_zhu2013layer}. The calculations based on this scheme
are proven to reproduce the measured spin polarization of photoelectrons
from Bi$_2$Se$_3$ quite well~\cite{S_zhu2014photoelectron,S_sanchez2014photoemission}.

We set $\phi_\alpha(\mathbf{r})=c_\alpha\, e^{-r^2/R_0^2}$, where the parameter $R_0$ is set to be $0.3$ times the lattice parameter, which is similar to the relevant atomic radii
and $c_\alpha$ is an atomic-type-dependent constant.
We checked that using the atomic valence $s$ orbitals as $\phi(\mathbf{r})$'s yields essentially the same results. For the final states of Bi$_2$Se$_3$, we set $c_\alpha$ for Se to be twice as large as that for Bi in order to match the atomic cross sections of the two elements~\cite{S_zhu2013layer}.


\begin{figure*}
\includegraphics[width=2\columnwidth]{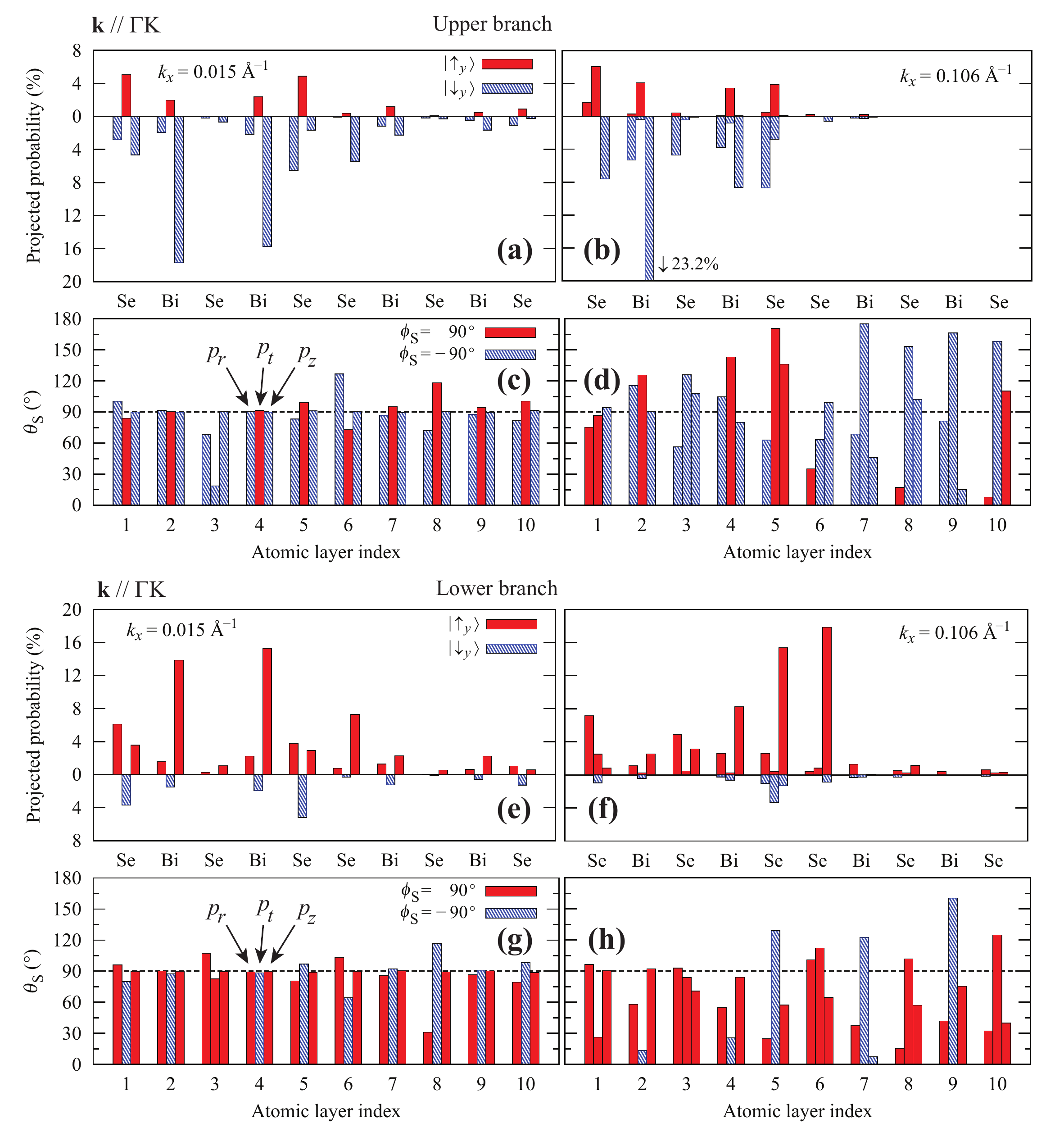}
\caption{(a) and (b) Projected probability of the surface states in the upper branch
with {\bf k} along $\mathrm{\Gamma K}$ and
at $k = 0.015\,\mbox{\AA}^{-1}$  [(a)] and at $k = 0.106\,\mbox{\AA}^{-1}$ [(b)].
The spin is quantized along $y$ [the tangential direction; see Fig.~1(a) of the main manuscript].
Atomic layer~1 is the topmost surface layer.
(c) and (d) The direction of the spin polarization corresponding to a spinor that
couples to each $p$ orbital, for the same surface states as in (a) and (b), respectively.
Spin direction $(\sin \theta_{\textrm{S}} \cos \phi_{\textrm{S}},\, \sin \theta_{\textrm{S}} \sin \phi_{\textrm{S}},\, \cos \theta_{\textrm{S}})$ is represented by the two angles $\theta_{\textrm{S}}$ and $\phi_{\textrm{S}}$.
(e)-(h) Similar quantities as in (a)-(d) for the lower branch.}
\end{figure*}

\end{document}